# Diffusion in the presence ofcorrelated dynamical disorder and coherent exciton transferin the non-Markovian limit

Rajesh Dutta[1] and Biman Bagchi[1,*]

[1]SSCU, Indian Institute of Science, Bangalore 560012, India.

*Email: bbagchi@iisc.ac.in

*Abstract*

*The presence of static off-diagonal disorder promotes coherent exciton transport while diffusive motion can be recovered in the presence of fluctuations in the diagonal and off-diagonal elements of the Hamiltonian. Here we studythe crossover induced by correlated dynamical disorder.We uncover a novel role of the excited bath states (ExBS) in dictating quantum coherence and quantum transport in dissipative quantum systems interacting with correlated bath.We solve both analytically and numerically the temperature dependent Quantum Stochastic Liouville equation (TD-QSLE) to study temperature dependence of quantum coherence in both linear chains and cyclic trimer (first three subunits of Fenna-Matthews-Olson(FMO) and also heptamer) complexes, using Haken-Strobl-Reineker Hamiltonian. In the non-Markovian limit where the lowering of temperature induces long-lasting quantum coherences, ExBSnot only determines the lifetime of coherences but also dictates the long time population distribution. We find a parallelism between classical and quantum systemsthrough transitions among excited bath states that provides a deeper insight about role of temperature inequilibrium distribution.The effects ofdynamic disorder and excited bath stateon quantum entanglement (through the calculation of concurrence) in single exciton manifold are demonstrated.*

The dynamics of a quantum system is determined by the degree of quantum coherence which in turn is determined by the off-diagonal elements of the density matrix. Coherent transport dominates when the static (or, the average) part of the off-diagonal coupling (J) is larger than the cumulative effects of fluctuations, determined by both the amplitude and the rate of fluctuations. Coherence can be destroyed by fluctuations due to the bath. In an extended system with dynamic disorder, the off-diagonal matrix element contains spatial variables, and averaging over the bath states is tricky. [1-3]Inclassical systems, the evolution of an initial non-equilibrium state towards equilibrium distribution is intimately connected with the corresponding, energy conserving transitions in the bath state [4]. The concept of temperature enters through detailed balance. The latter condition leads, in the long time, to the Boltzmann distribution $\frac{N_j}{N_i} = e^{-\beta \Delta E_{ij}}$, $i$ and $j$ are the two states of the system. Theinverse of temperature, $\beta=1/k_B T$, is defined, interestingly, in terms ofenergy gap between the bath states and a ratio of the occupation number in different bath states.The classical derivation cannot be used in quantum mechanics as a rate equation approach is not valid, due to the presence of quantum coherence. Second major difficulty lies with the inclusion of temperature inquantummechanical description. The third issue is thatthe distribution itself is non-Boltzmann at low temperatures.Effects of temperature on thetime evolution of an initial non-equilibrium state of a quantum system, particularly when the system is extended in space, poses formidable difficulty. It is well-established that increase in the amplitude of static disorder causes a localization transition in two- and three-dimensional systems and system becomes diffusionless [1-3]. There has been less study of the same In the presence of dynamical disorder (when coupling elements fluctuate with time).

RecentlyFleming and coworkers studied the effects of temperature on quantum coherence in photosynthetic complexes [5-7].A theoretical study of the same was carried out on the transport processes in conjugated polymers [8-10].A temperature dependent study to explore the role of coherence and of the bath states, in establishing the equilibrium distribution (even at high temperatures) does not seem to exist.

In this we work we study the interplay between temperature and quantum coherence in giving rise to unusual time dependence in the approach of the system to its equilibrium distribution which is clearly non-Boltzmanat low temperature [41, 12]. We find that excited bath states play important role, somewhat similar to their role in establishing equilibrium distribution in classical systems. We have demonstrated this unique role in excited bath states in three different systems, each with unique features.

### A. Excited bath states in photosynthesis

We first consideratrimer system that consists of the first three sites of FMO complex so that we can investigate the propagation of coherence in the presence of spatially and temporally correlated bath. We also explore the role of fluctuation strength and bath correlation time in the evolution of initial non-equilibrium state towards equilibrium distribution.

We employ the well-known Haken-Strobl-Reineker-Silbey (HSRS) exciton Hamiltonian, [13-15]

$$H_{tot} = H_S + H_B + H_{int} \tag{1}$$

where system (exciton) Hamiltonian is defined as

$$H_S = \sum_k E_k |k\rangle\langle k| + \sum_{\substack{k,l \\ k \neq l}} J_{kl} |k\rangle\langle l| \tag{2}$$

where $E_k$ is the energy of an electronic exciton localized at site $k$ and $J_{kl}$ is the time-independent off-diagonal interaction between excitations at site $k$ and $l$. We assume that the bath is a collection of harmonic oscillators

$$H_B = \sum_j \left( \frac{p_j^2}{2m_j} + \frac{1}{2} m_j \omega_j^2 x_j^2 \right) \tag{3}$$

If the system-bath interaction Hamiltonian is assumed to be given as $H_{SB} = -VX$ where, V consists of system part and X is a collective bath variable, $X = \sum_j c_j x_j$, then one can use Feynman-Vernon influence functional [16] to eliminate X from the total Hamiltonian to write it as

$$H_{tot} = H_S + V(t) \tag{4}$$

Feynman-Vernonallows for a procedure to obtain V(t). The procedure is the same as Kubo'stheory of reduction in degrees of freedom embodied in Kubo-Mori-Zwanzig (KMZ) approach to time-dependent statistical mechanics.When bath is a collection of harmonic oscillators and the coupling is linear, V(t) can be approximated as a Gaussian random variable under rather general conditions.

A joint probability distribution is defined in system and bath variables as follows

$$P(\rho, V, t) = \langle \delta(\rho - \rho(t)) \delta(V - V(t)) \rangle. \tag{5}$$

We can adopt the density matrix formalism from quantum Liouville equation to write the quantum stochastic Liouville equation in two variables as

$$\frac{\partial}{\partial t}P(\rho,V,t) = \frac{i}{\hbar}\frac{\partial}{\partial \rho}[H_{tot},\rho]P + \Gamma_V P \qquad (6)$$

where $\Gamma_V$ is a stochastic diffusion operator. For Gaussian bath it is a Fokker-Planck operator [17].

**(SM)**

We next define a reduced QSLE by averaging the density matrix as,

$$\sigma(t) = \int d\rho\, \rho\, P(\rho,V,t) \qquad (7)$$

By recombination of Eq. (6) and (7) and followed by integration by parts, the QSLE [18, 19] for the full density matrix reduces to

$$\frac{\partial \sigma}{\partial t} = -\frac{i}{\hbar}[H(t),\sigma] + \Gamma_V \sigma \qquad (8)$$

The temperature corrected QSLE was derived by Tanimura and Kubo [20] using dynamical approach. The equation of motion (EOM) in reduced density matrix can be given as,

$$\frac{\partial \sigma(V,t)}{\partial t} = \left[-\frac{i}{\hbar}H(t)^x - b\frac{\partial}{\partial V}\left(V + \frac{\partial}{\partial V}\right) - \frac{i\beta b}{2}\left(V + \frac{\partial}{\partial V}\right)V^o\right]\sigma(V,t) \qquad (9)$$

The highly non-trivial nature of Eq. (9) precludes an easy analytical solution but one can follow Kubo's method of expansion of the RDM in the eigen-states $|b_m\rangle$ of the bath operator $\Gamma_V$ with $\sigma_m$ as the expansion coefficient

$$\sigma = \sum_m \sigma_m |b_m\rangle \qquad (10)$$

This leads to a hierarchical equation of motion as,

$$\frac{\partial \sigma_m}{\partial t} = -\frac{i}{\hbar} H_{ex}^x \sigma_m - \frac{iV}{\hbar} V^x (\sigma_{m+1} + \sigma_{m-1}) - \frac{\beta Vb}{2} V^o - mb\sigma_m \qquad (11)$$

TheEOM involves three termsrepresenting the effect of the bath. (i) Secondterm gives the linear coupling betweensystem and different bath states, (ii) third term due to the temperature dependence of the bath modes, and (iii) the last term that gives rise to straightforward decay of the bath states.

Note that $\sigma_0$ is the equilibrium bath state, signaling that the bath itself is in equilibrium, with no decay (m=0) in Eq.11. The system-bath interaction occurs through the excited bath states (m =1,2,3 ….). The connection with classical systems is intriguing. [4]

For further progress, the stochastic bath perturbation HamiltonianV(t)is next decomposed into diagonal and off- diagonal fluctuations

$$V(t) = \sum_k |k\rangle\langle k| V_d^{(k)}(t) + \sum_{\substack{k,l \\ k \neq l}} |k\rangle\langle l| V_{od}^{(kl)}(t) \qquad (12)$$

Here, $V_d^{(k)}(t)$ and $V_{od}^{(kl)}(t)$ denote diagonal (site energy at *k*) and off-diagonal (connecting sites *k* and *l*) parts of the fluctuating perturbation $V(t)$ (each with average zero). Different models for the space dependence of these functions are used. (i) Spatially correlated (same) bath model (ii)The uncorrelated (independent) bath model. We *use "C" and "UC"assuper-scripts to denote correlated and uncorrelated bath respectively.*

We assume *V(t)*to be a Gaussian Markov process [21] which is realistic for harmonic oscillator bath. We also employ atwo-statePoisson stochastic model process [22-24].*Our model is quite* *similar to the two-level dissipative system considered by Leggett and co-workers*[25]. The system Hamiltonian which considers diagonal site energies and off-diagonal inter-site coupling can be reduced to Spin-Boson Hamiltonian [26]. However, most of these calculations are limited to a Markovian description of the stochastic perturbation. *Our assertion is that such a Markovian description underestimates the effects of coherence.*

Recently the role of fluctuating environment is explored in excitation energy transfer in photosynthetic complex. Fluctuations not only destroy the coherence but also it can facilitate the coherence [27, 28].In this work we quantify the role of fluctuation strength (V) and rate of fluctuation (b) in propagation of coherences. We also explore quantum entanglement in single exciton manifold in different memory regime where temperature effect is crucial.

We now proceed to explicitnumerical calculation of the cyclic hetero trimer. Here we use system parameters from FMO [29] Hamiltonian(**SM**) [17].We have numerically solved the coupled EOM using Runge-Kutta fourth order method. Coherence can be represented for uncorrelated bath model as follows,

$$Coherence(m,n) = \langle m | \sigma_{N \atop \prod_{i=1}^{} a_i} | n \rangle \tag{13}$$

where, m and n are the site number, N is the total number of sites and $a_1, a_2, a_3, .... a_N$ all the elements could be zero or onedepending upon whether it isin equilibrium or excited states of Poisson bath. For Gaussian bath $a_i$ can take values from 0 to $\infty$. There couldbe small changes in amplitude of oscillation after 4$^{th}$ bath states but the nature of coherences will remain thesame. For this reason, we have considered upto 4$^{th}$ bath states for both temperature dependent and independent case.

The correlated bath model is not fully applicable to real photosynthetic system and we thus study spatially uncorrelated bath model to calculate coherences in different bath states. We have calculated coherence analytically for the dimer system with only off-diagonal dynamic disorder[17]. (**SM**) Role of temperature correction term can be easily understood from these expressions. Numerically, we evaluate coherences from both temperature dependent and independent QSLE for uncorrelated bath model. At high temperature limit these two merge into each other.

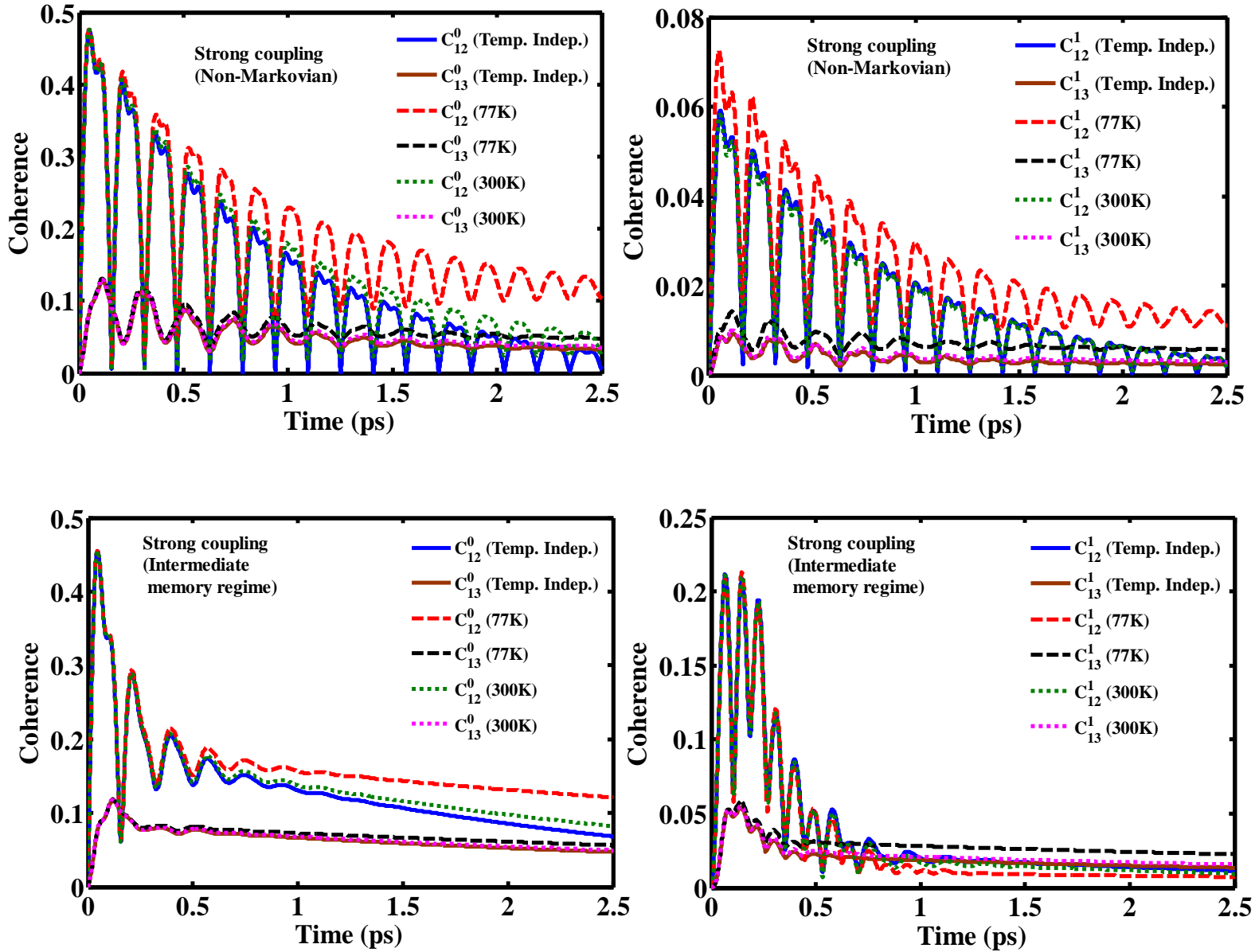

**Figure1.(a)-(b) depict absolute value of coherencein equilibrium bath state and excited bath states for trimer system at $V_d^{UC}$ = 50 cm$^{-1}$ and $(b_d^{UC})^{-1}$ = 10 fs.(c)-(d) indicates absolute value of coherence in equilibrium bath state and excited bath states for trimer system at $V_d^{UC}$ = 50 cm$^{-1}$ and $(b_d^{UC})^{-1}$ = 250 fs.Solid line indicates results obtained from temperature independent calculation. Dashed line designates absolute value of coherence at 77K. Dotted line indicates absolute value of coherence at 300K.0 and 1 indicate equilibrium and excited bath.**

In **Figure 1** the temperature dependence of absolute values of coherencesis plotted indifferent regimes. In the non-Markovian limit the decay of coherence is slow and the dynamics is oscillatory. *With decrease in temperature, the decay further slows down for both coherence in EBS and ExBS. However, the effect of temperature is more prominent for coherence between nearest neighboring sites than non-neighbors.* Time scales for decay of coherence in EBS and ExBSare more or less the same, although the contribution towards excitation transfer dynamics (ETD) is largerby theEBS. In strong coupling but intermediate memory regime, the effect of temperature is only observed for the coherence between the nearest neighbors in EBS. However, the contribution towards ETD is quite comparable for both coherence in EBS and ExBS.

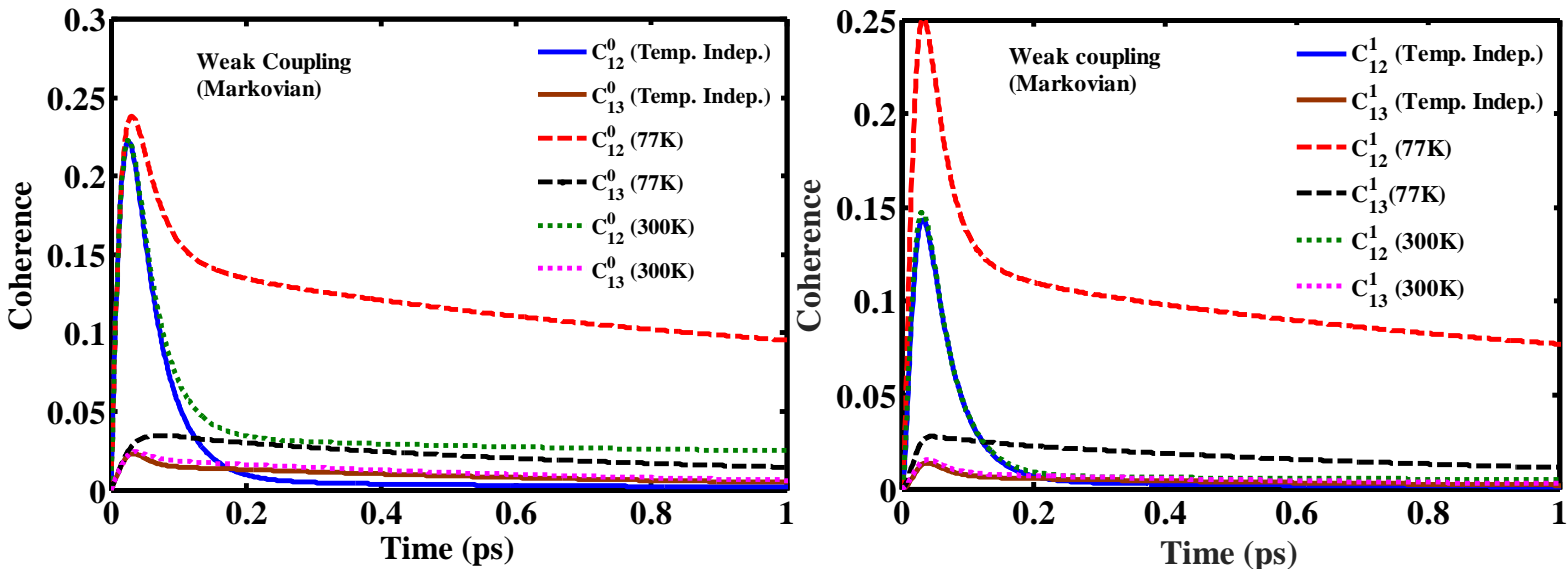

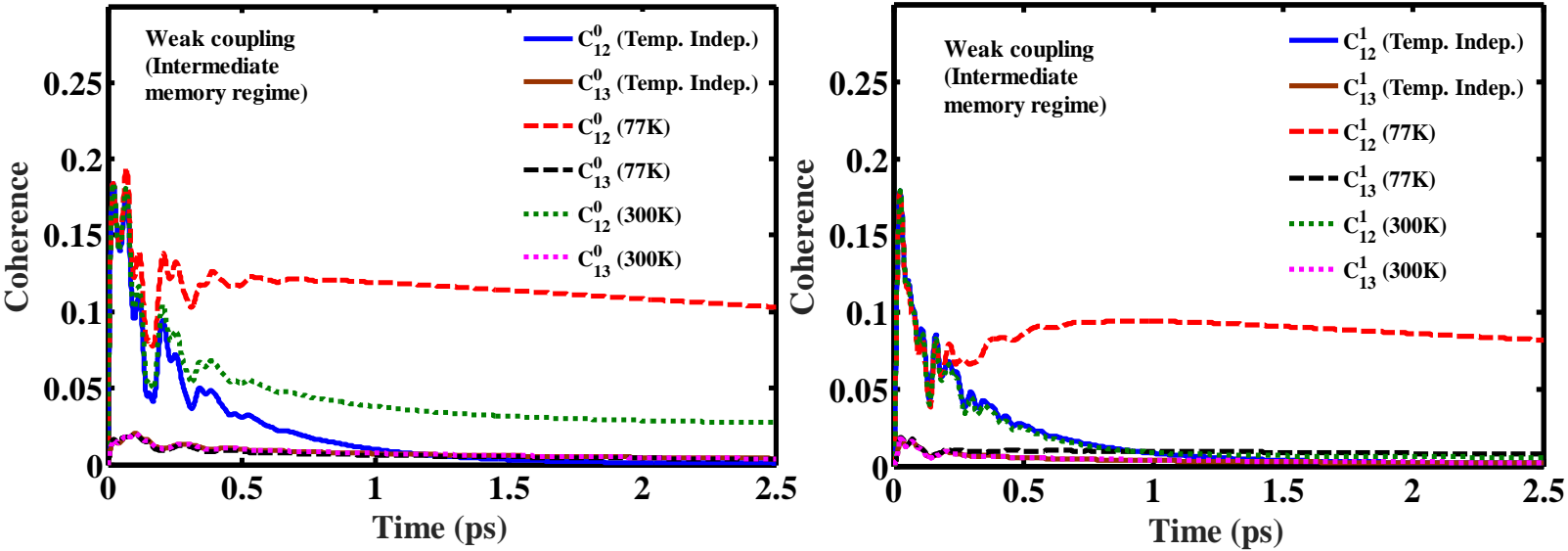

**Figure2.(a)-(b) depict absolute value of coherence in equilibrium bath state and excited bath states for trimer system at $V_d^{UC}$ = 350 cm$^{-1}$ and $\left(b_d^{UC}\right)^{-1}$ = 10 fs.Figures (c)-(d) depict absolute value of coherence in equilibrium bath states and excited bath states for trimer system at $V_d^{UC}$ = 350 cm$^{-1}$ and $\left(b_d^{UC}\right)^{-1}$ = 250 fs.Solid line indicates results obtained from temperature independent calculation. Dashed line designates absolute value of coherence at 77K. Dotted line indicates absolute value of coherence at 300K.0 and 1 indicate equilibrium and excited bath. At high temperature limit temperature independent and dependent case show similar behavior.**

In**Figure 2**we plot the absolute value of coherence in weak coupling-Markovian and weak coupling intermediate regime. For both the cases, we observe pronounced temperature effect in coherencebetweennearest neighbor sites. However, coherence in EBS and ExBS contribute similarly in ETD. In weak coupling-Markovian limit we observe non-oscillatory decay of coherence. We also observe that with increase in the number of ExBS the contribution of coherences towards the ETD decreases [17]. (**SM**)

Time scale of oscillation of coherencesis determinedby the competitive dependence on the parameters, *V, b, $V^2/b$* and *J*. Increase in temperature helps in transition from coherent to incoherent state and vice versa. *Lowering temperature helps in preserving phase relation between excitonic states for long time.* In the long time limit when oscillation disappears, low temperature helps in slow decay which essentially indicates localization of the energies on corresponding site.

Coherences between non-local sites create interference between pathways of energy transfer. Non-local coherences open up efficient channels for energy transfer and facilitate energy transfer dynamics. For our model system the non-local coherence leads to energy transfer from site 1 to site 3 by avoiding the barrier i.e. site 2.

Whenfluctuation strength (V) and rate (b) are large and the ratio $V^2/b$ is greater than 2J, the oscillations vanish.In this limit for uncorrelated bath case,coherence in ExBS decaysfaster than coherence in EBS. However, in intermediate limit that is appropriate for photosynthetic and conjugated systems, coherence in ExBS and coherence in EBS play equally important role.

*Concurrence* [30] *is often used as ameasure of quantum entanglement in single exciton manifold.*For bipartiate entanglement between two chromophores, it is defined as the double of the absolute value of coherence [31].At short time the increase in entanglement leads to quick delocalization of excitation. Considerable amount of entanglement is observed even for non-nearest neighbors in strong coupling limit and for EBS.*Temperature independent QSLE provides an equal population of all sites in long time limit which indicate the vanishing of quantum entanglement in long time limit*. However, temperature dependent QSLE shows that unequal population leads *to non-vanishing quantum entanglement at long time limit.*

## B. Role of excited bath states in restoring long time equilibrium distribution

Here we discuss a novel role of the excited bath states in establishing the long time equilibrium distribution. Fortunately, the analysis can be carried out analytically because we can solve the dimer problem exactly for Poisson bath for all the matrix elements.

As shown in the [17] (**SM**), the long time limit of population of the two levels is given by :

$$\text{Population of site1 } \langle 1|\sigma_0|1\rangle^C_{t\to\infty} = \frac{1}{2} - \frac{\beta\Delta}{4} \tag{14}$$

$$\text{Population of site2 } \langle 2|\sigma_0|2\rangle^C_{t\to\infty} = \frac{1}{2} + \frac{\beta\Delta}{4} \tag{15}$$

$$\frac{P_1}{P_2} = \frac{\langle 1|\sigma_0|1\rangle^C_{t\to\infty}}{\langle 2|\sigma_0|2\rangle^C_{t\to\infty}} = \frac{\frac{1}{2} - \frac{\beta\Delta}{4}}{\frac{1}{2} + \frac{\beta\Delta}{4}} \approx \frac{e^{-\frac{\beta\Delta}{2}}}{e^{\frac{\beta\Delta}{2}}} = e^{-\beta\Delta} \tag{16}$$

where, $\Delta = E_1 - E_2$.

From the above expression, it is clear that in the long time limit populations obey Boltzmann distribution to the first order in β. Above derivation is strictly valid in the limit of high temperatures. Coherences have interesting limiting expressions in the long time limit as

$$\left|\langle 1|\sigma_0|2\rangle\right|^c_{t\to\infty} = \frac{J}{2k_BT}$$

$$\left|\langle 1|\sigma_1|2\rangle\right|^c_{t\to\infty} = \frac{V^c_{od}}{2k_BT} \tag{17}$$

These are indeed interesting as they show that while coherence in EBS is given by *J*, the coherence in ExBS is given by $V^c_{od.}$

We now show that these long time limits are completely different for the temperature independent case. In this case, the term *d* in the coupled EOM will be zero.

Hence in long time limit, we obtain

Population of site 1=Population of site 2=$\langle 1|\sigma_0|1\rangle^C_{t\to\infty} = \langle 2|\sigma_0|2\rangle^C_{t\to\infty} = \frac{1}{2}$ (18)

We can also show that in the long time limit, the coherences go to zero,

$$\left|\langle 1|\sigma_0|2\rangle\right|^C_{t\to\infty} = \left|\langle 1|\sigma_1|2\rangle\right|^C_{t\to\infty} = 0 \qquad (19)$$

For temperature dependent case, populations asymptotically reach Boltzmann distribution. *However, this is seen to be intimately connected to the presence of non-zero value of the coherence in ExBS.* Coherence in the EBS is only dictated by inter-site coupling J, whereas, coherence in the ExBS is only governed by fluctuation strength in long time limit.

Numerically we also findthat for uncorrelated bath cases at high temperature the populations obtained from temperature dependent QSLE are exactly the same as the Boltzmann distribution [17]. (**SM**)

We find that the population difference and non-vanishing coherence are connected through the following interesting relation

$$\left[\langle 1|\sigma_0^{exc}|1\rangle_{(t\to\infty)} - \langle 2|\sigma_0^{exc}|2\rangle_{(t\to\infty)}\right] = 2\cos ec2\theta \ \langle 1|\sigma^{site}|2\rangle_{(t\to\infty)} \qquad (20)$$

where, $\tan 2\theta = \frac{2J}{\Delta}$.

Eq. (20) shows the off-diagonal elements in site basis do not vanish in the long time limit, as the populations of different energy levels in eigen-basis are not equal [32]. As we have shown above, the population approaches Boltzmann distribution at high temperature, Eq. (20) provides an important correlation between coherences and Boltzmann distribution.

## C. Role of excited bath states in quantum transport

It is well known that static disorder can localize exciton (Anderson localization) and that dynamic disorder can allow the transport to survive [33]. We now consider a one dimensional chain of regularly placed two-level systems [34]. This is a model studied extensively by Haken and coworkers and also by Silbey. In the Markovian limit of bath fluctuations, Haken and Reinekerderived the following expression for the exciton diffusion co-efficient

$$D = a^2 \left[ 2\gamma_{od} + \frac{J^2}{\gamma_d + 2\gamma_{od}} \right] \quad (21)$$

where $\lim\limits_{V_d, b_d \to \infty} \frac{V_d^2}{b_d} = \gamma_d$ and $\lim\limits_{V_{od}, b_{od} \to \infty} \frac{V_{od}^2}{b_{od}} = \gamma_{od}$

Eq.21 shows an intricate dependence of diffusion on the rate parameters $\gamma_{od}$ and $\gamma_d$. Perhaps the most important is the limit where the off-diagonal rate goes to infinity, keeping the diagonal rate fixed.

However, the above expression is strictly valid in the Markovian limit. In the non-Markovian limit, it is extremely hard to obtain an analytical expression. This is precisely because of the coupling of the equilibrium bath states to excited bath states.

One can derive a somewhat more general expression than that of Eq.21 valid in the non-Markovian limit in the absence of off-diagonal dynamic disorder and $b_d \to 0$ as follows

$$D = \frac{a^2 J^2}{5 b_d} \quad (22)$$

From Eq. (22) it is clear that diffusion co-efficientdiverges at low $b_d$ value and at high $b_d$ value diffusion co-efficient is proportional to $b_d$. This essentially indicates that decrease in the $b_d$ value increases the coherence and exciton transport is also coherent. In the same limit one can also obtain fully coherent motion for correlated bath case.

We show analytically that in the absence of static disorder or SHE, coherence is propagated through equilibrium bath states in Markovian limit. However, in non-Markovian limit coherence in EBS and ExBS plays equally important role. Hence one can disentangle the role of coherence in EBS and ExBS in the localization-delocalization process where the memory effect is also crucial.

However, in the absence of dynamic diagonal disorder, localization cannot be achieved from dynamic off-diagonal disorder.It helps the delocalization of exciton.

Below we summarize the main features of our study.

(1) In the non-Markovian limit we observe coherent dynamics and transition from coherent to incoherent dynamics while going from non-Markovian to Markovian limit. *We analytically show for correlated bath model coherence is propagated through ExBS which only contains non-vanishing term even in long time limit in absence of site energy heterogeneity.* In presence of difference in site energy and in the long time limit, coherences in EBS and ExBSare governed by off-diagonal coupling and fluctuation strength respectively. *In strong coupling and non-Markovian limit coherence in EBS is more dominant than coherence in ExBS. However, time scale of decay is more or less the same.*

(2) Non-local coherences between non-neighboring sites lead to creation of new pathways of energy transfer. Non-local coherences help overcome the energy barrierthus facilitating energy transfer dynamics. Double of the absolute value of coherence is known as concurrence. This is an entanglement measure for single exciton manifold. *Temperature* *dependent study shows non-vanishing quantum entanglement even in long time limit*. No such non-vanishing entanglement is observed in case of temperature independent case.

(3) We establish a quantitative relation *between temperature dependent equilibrium distribution and quantum coherences*. For both correlated and uncorrelated bath case, non-zero coherences even in long time limit signify finite phase relation between the states even in the long time. In long time limit coherence is proportional to the population differences thus leading to Boltzmann distribution asymptotically.

(4) For the correlated bath case, we show analytically that coherence in excited bath decays faster in the Markovian limit than that of the equilibrium bath states [17] (**SM**). Hence, in the absence of site energy heterogeneity or static disorder, we find that localization becomes correlated with the coherence in equilibrium bath states.

(5) We observe coherences for temperature dependent/independent case and also in the presence/absence of site energy heterogeneityfor uncorrelated bath. We obtain a relation between coherences and localization-delocalization. In the absence of static disorder or site energy heterogeneity, coherence in long time limit propagates only through ExBS. However, in the presence of static disorder or site energy heterogeneity,coherences in both EBS and ExBS show non-vanishing behavior in thelong time limit. This ensures the significant correlation between localization-delocalization and coherences in EBS and ExBS.

In a series of papers, Cao et. al. [11, 12] employed perturbative approach to obtain the steady state distribution. Although they did not obtainlow temperature distribution, the high temperatureresult is quite similar to our result. From simulation they have calculated the Bloch sphere rotation and concluded that the steady state distribution is canonical at high temperature limit as well as for large value of system-bath coupling parameter. However, the situation deviates largely at small coupling limit as well as at low temperature limit where the steady state distribution is non-canonical.

The existence and decay of quantum coherences in extended systems can also be studied via line shapes [35] as a direct observation in time domain is not always feasible. Line shape is defined as the Fourier transform of transition dipole moment correlation function averaged over EBS. Analysis of the line shape equation shows thatall the bath states are involved. However, faster decay of higher bath states, at least in the Markovian limit, makes it possible to theoretically disentangle the effects of coherence.

The static modulation limit of our uncorrelated bath case of our extended model approaches the Anderson model of localization of wave function.As the rate of fluctuation *(b)*increases, we might observe an intermediate situation where fluctuation induces delocalization. The localization should be restored in the fast modulation limit. This deserves further study.

## Acknowledgements

We thank Prof. J. Cao (MIT) and Prof. Y. Tanimura (Kyoto) for several useful discussions about steady state population distribution. We acknowledge the Department of Science and Technology (DST, India) for partial support of this work. B. Bagchi thanks Sir J. C. Bose Fellowship for providing partial financial support. R. Dutta thanks Dr. S. Sarkar, Mr. SayantanMondal, Ms. G. Shivali, Mr. S. Kumar and Ms. S. Mondal for carefully going through the manuscript.